# On the topological surface states of the intrinsic magnetic topological insulator Mn-Bi-Te family

Yuan Wang[1], Xiao-Ming Ma[1], Zhanyang Hao[1], Yongqing Cai[1], Hongtao Rong[1], Fayuan Zhang[1], Weizhao Chen[1], Chengcheng Zhang[1], Junhao Lin[1], Yue Zhao[1#], Chang Liu[1#], Qihang Liu[1#], and Chaoyu Chen[1#]

[1] Shenzhen Institute for Quantum Science and Engineering (SIQSE) and Department of Physics, Southern University of Science and Technology (SUSTech), Shenzhen 518055, China.

[#]Correspondence should be addressed to Y.Z. (zhaoy@sustech.edu.cn), C.L. (liuc@sustech.edu.cn), Q.L. (liuqh@sustech.edu.cn) and C.C. (chency@sustech.edu.cn)

**ABSTRACT**

We review recent progress in the electronic structure study of intrinsic magnetic topological insulators (MnBi$_2$Te$_4$)•(Bi$_2$Te$_3$)$_n$ ($n = 0, 1, 2, 3$) family. Specifically, we focus on the ubiquitously (nearly) gapless behavior of topological surface state Dirac cone observed by photoemission spectroscopy, even though a large Dirac gap is expected because of surface ferromagnetic order. The dichotomy between experiment and theory concerning this gap behavior is perhaps the most critical and puzzling question in this frontier. We discuss various proposals accounting for the lack of magnetic effect on the topological surface state Dirac cone, which are mainly categorized into two pictures, magnetic reconfiguration, and topological surface state redistribution. Band engineering towards opening a magnetic gap of topological surface states provides great opportunities to realize quantized topological transport and axion electrodynamics at higher temperatures.

1.    **Introduction**

Magnetism has been used and studied over millennia. Yet, this branch of physics keeps flourishing in recent years with emerging states of magnetic matters such as quantum spin liquid [1,2] and two-dimensional (2D) magnets [3,4]. By comparison, the first two decades of the new millennium embraced the triumph of topological states of matter, which has revolutionized our knowledge of crystalline materials by introducing topological invariants to categorize their electronic structure [5,6]. In 2007, the realization of the quantum spin Hall effect (QSHE) based on a 2D HgTe/CdTe quantum well [7,8] opened a new era of exploring topological phases and materials in condensed matter. Electronically, this QSH state is insulating with a bulk gap separating the conduction and valence bands but has a pair of one-dimensional (1D) conducting edge states. These 1D topological edge states are protected by time-reversal symmetry, wherein elastic backscattering by nonmagnetic impurities is forbidden, holding potential for dissipationless spintronics. The QSHE state can be classified by a type of topological invariant called $Z_2$ invariant [9] and is now recognized as the first example of 2D time-reversal invariant topological insulator (TI) with $Z_2 = 1$. $Z_2$ classification can be generalized to three-dimensional (3D) to describe "weak" and "strong" 3D TIs [10]. A 3D strong TI has ubiquitous, gapless topological surface state (TSS) with helical spin texture due to the spin-momentum locking. In 2008, 3D TI was first realized based on Bi-Sb alloys [11] with multiple TSSs crossing the Fermi level 5 times. Up to now, there have been hundreds of materials predicted as 3D strong TI [12-14] and dozens of them have been experimentally verified, usually through direct observation of their TSS Dirac cone by angle-resolved photoemission spectroscopy (ARPES) [15-17]. Among them, the most representing one is $Bi_2Se_3$ family of materials [18-21] found in 2009. $Bi_2Se_3$ family is now considered as the "hydrogen atom" of topological materials due to its simple and elegant electronic structure. It is a semiconductor with a bulk gap of $\sim 0.3\ eV$, among the few with sizable bulk gap to manifest surface state transport. It has only one TSS Dirac cone at the Brillouin zone (BZ) center, with the Dirac point located close to the middle of bulk gap. It consists of -Se-Bi-Se-Bi-Se- quintuple layers (QLs) stacking along $c$ axis and bound by van der Waals (vdW) interaction, easy for device fabrication and epitaxial growth. Importantly, it is chemically stable and the robustness of TSS when exposed to ambient environment has been firmly established [22]. These excellent properties distinguish $Bi_2Se_3$ family from others for the exploration of novel topological effects. Particularly, the quantum anomalous Hall effect (QAHE), a time-reversal-symmetry-breaking version of QSHE, was predicted in 2010 [23] and then realized in 2013 [24] based on magnetically doped films of this family (Fig. 1).

Our story of topological states of matter has reached a point where this rising star meets a classic field of physics, magnetism. However, this is not their first rendezvous. Back in 1980, the quantum Hall effect (QHE) was observed from a 2D electron gas system subjected to a strong magnetic field [25]. This phenomenon was later explained theoretically as a topological property of the occupied bands in BZ [26]. In the current context, we can call the QHE the first topological insulator ever found, classified by a time-reversal-symmetry-breaking topological invariant called Chern number $C$ (originally the well-known Thouless–Kohmoto–Nightingale–Nijs invariant [26]). The applied magnetic field in QHE can be replaced by the intrinsic magnetization of the material, leading to QAHE. In this regard, QHE and QAHE states are both Chern insulators. In 2D Chern insulators, 1D gapless edge states emerge at the boundary between the Chern insulator and ordinary insulator (vacuum) because of the distinct band topology. Like the 1D helical edge states of QSHE, the 1D chiral edge state in QAHE can propagate along one direction with forbidden backscattering, suitable for developing low-power-consumption electronics without the need for applied magnetic field.

The prediction and realization of QAHE [23,24] represent a breakthrough in fundamental physics, yet much effort is needed toward its practical application. On the one hand, Hall bar devices based on magnetically doped films require sophisticated epitaxial growth and microfabrication procedures that only a few labs in the world can accomplish [24,27-34]. On the other hand, atomic doping brings disorders to the crystal and inhomogeneity to the electronic structure, resulting in much reduced effective exchange gap of the TSS, where the realization temperature was as low as $30\ mK$ at first [24] and the current record is $2\ K$ [30] after ten years' effort. In this context, intrinsic magnetic TI, which combines magnetic order and band topology in the same material without the need of doping, is highly desired.

In a magnetic TI, while the time-reversal symmetry $\Theta$ is broken by the magnetic order, its combination with certain magnetic lattice symmetry such as rotation $C_n$ and fractional translation $T_{1/2}$ can retain an equivalent time-reversal symmetry and the system can still be classified by the topological $Z_2$ invariant. This was first discussed in the theoretical proposal of antiferromagnetic TI (AFM TI) in 2010 [35]. In an AFM TI, both $\Theta$ and $T_{1/2}$ are broken but the combination $S = \Theta T_{1/2}$ is preserved, leading to a topologically nontrivial phase which shares with 3D strong TI the topological $Z_2$ invariant and quantized magnetoelectric effect. The difference is that, while 3D TIs have symmetry protected gapless TSS at all surfaces, 3D AFM TIs have intrinsically gapped TSS at certain surfaces with broken $S$ symmetry. The gapped TSSs in 3D AFM TIs carry a half-quantized Hall conductivity ( $\sigma_{xy} = e^2/2h$ ), which may aid experimental confirmation of quantized magnetoelectric coupling (Fig. 1). Although with fascinating properties, the material realization of an intrinsic AFM TI was initiated until 2017. First-principles calculation proposed that by inserting MnTe bilayer into the quintuple layer of Bi2Te3, the septuple layers of MnBi$_2$Te$_4$ host a robust QAH state [36,37]. The material was first experimentally realized with thin films via molecular beam epitaxy [38]. Theoretical works on single crystalline MnBi$_2$Te$_4$ were reported in 2019 and revealed its fertile topological states of matter [39-42]. Since the successful preparation of single crystal MnBi$_2$Te$_4$, the surge of intrinsic magnetic TI based on MnBi$_2$Te$_4$ family of materials started.

Now we know more details concerning the structural, magnetic, and topological aspects of MnBi$_2$Te$_4$ AFM TI. Its $R\bar{3}m$ lattice consists of layered -Te-Bi-Te-Mn-Te-Bi-Te- units (SL) stacking along the $c$ axis and the layers are bounded by vdW force [43,44]. In the ground state below $T_N \sim 24.6\ K$, Mn ions ($S = 5/2$ of 2 + valence) with a large magnetic moment of $\sim 5\ \mu_B$ form a ferromagnetic (FM) layer with moments pointing out-of-plane. These FM layers couple each other in an AFM way along the $c$ axis [45,46] (A-type AFM). The bulk gap is around $220\ meV$ from calculation [42] but only $\sim 130\ meV$ as directly observed by ARPES [47-49]. At the natural cleavage plane (0001), the TSS gap of $\sim 88\ meV$ at the Dirac point is expected due to the $S$ breaking [39-42]. This sizable TSS gap is the key ingredient enabling the observation of exotic phenomena such as quantized magnetoelectric coupling [35,50,51], axion electrodynamics [52,53], QAHE [24,27-34,54] and chiral Majorana fermions [55-58] at much higher temperatures. In fact, quantum transport experiments have revealed the existence of a 2D Chern insulator with QAHE observed at 1.4 $K$ based on 5 SLs [59], and the characteristics of an axion insulator state based on 6 SLs [60], both at zero magnetic field. Under a perpendicular magnetic field (15 $T$), characteristics of high-Chern-number quantum Hall effect without Landau levels contributed by dissipationless chiral edge states are observed, indicatinga well-defined Chern insulator state with $C = 2$ (9, 10 SLs) [61]. Its intralayer FM and interlayer AFM configuration exhibits layer Hall effect in which electrons from the top and bottom layers deflect in opposite directions due to the layer-locked Berry curvature, resulting in the characteristic of the axion insulator state (6 SLs) [62]. In addition to these experimental observations, it is further shown by theory that manipulating its magnetic and structural configuration can give rise to many new topological states. For example, the flat Chern band in twisted bilayer $MnBi_2Te_4$ may boost fractional Chern insulator and $p + ip$ topological superconductor [63]; changing the stacking order between $MnBi_2Te_4$ SL and $Bi_2Te_3$ QL may lead to novel states such as QSHE insulator with and without time-reversal symmetry [64]; magnetic ground states other than A-type AFM may lead to different phases such as Weyl semimetal [65,66] and higher-order topological Möbius insulator [67]. These predictions (Fig. 1) certainly deserve further experimental efforts.

There have been several reviews/perspectives on this intrinsic magnetic TI family [68-72], with distinct emphases on theoretical, computational and transport study, respectively. In this review, we focus on the electronic structure of $(MnBi_2Te_4) \cdot (Bi_2Te_3)_n$ ($n = 0, 1, 2, 3$) family, which has long been a subject of much debate. We will first review the ARPES observation of ubiquitously (nearly) gapless behavior of TSS Dirac cone from $MnBi_2Te_4$ and $MnBi_2Te_4$ SL termination, as well as the band hybridization features from $Bi_2Te_3$ QL terminations. While the significantly reduced TSS gap size of $MnBi_2Te_4$ termination deviates from that obtained by first-principles calculations, it is not against the relatively low temperature for the observation of QAHE, suggesting that the effective magnetic moments for the TSS may be diminished. There are also experimental evidence suggesting the robust A-type AFM order at the topmost SL layers from magnetic force microscopy (MFM), polar reflective magnetic circular dichroism (RMCD) and X-ray magnetic circular/linear dichroism (XMCD/XMLD) measurements. Furthermore, the magnetic splitting of certain bulk quasi-2D bands seems to validate the effect of magnetic order on the low-energy band structure. Bearing these established experimental results in mind, we then discuss the validity of possible scenarios proposed to account for the (nearly) gapless TSS from $MnBi_2Te_4$ termination, such as surface magnetic reconstruction, TSS redistribution, defect and self-doping effects, etc. Future band engineering towards opening a magnetic gap at the TSS Dirac point via approaches such as magnetic manipulation, element substitution, chemical potential and material optimization is proposed, which would provide great opportunities to for the realization of QAHE and topological magnetoelectric effect at higher temperatures.

## 2. Ubiquitously gapless TSSs

Due to the $S$ breaking at the natural cleavage plane (0001), below $T_N \sim 24.6\ K$ AFM TI MnBi$_2$Te$_4$ is expected to show a magnetic gap $\sim 88\ meV$ at the Dirac point of TSS [39-42]. Earlier ARPES investigations on the single crystals reported gapped TSS behavior with a Dirac gap ranging from 70 $meV$ to 200 $meV$ [42,73,74], in line with the theoretical prediction. However, the photon-energy dependent gap size indicates its bulk nature rather than a surface origin. Indeed, our systematic photon-energy-dependent ARPES measurements show that the bulk gap separating the bulk valence and conduction bands varies from 130 meV to 200 meV from bulk BZ Γ to Z [47]. Astonishingly, the TSS Dirac cone remains gapless below or above $T_N \sim 24.6\ K$, as first reported by our group (data shown in Fig. 2a) and others [47-49]. Such observations of (nearly) gapless TSS Dirac cone on MnBi$_2$Te$_4$ (0001) surface below and above the AFM order temperature have been further repeated [75-79]. To date, there have been extensive efforts to explain the origin of this gapless behavior, which will be reviewed in the following sections.

The striking violation against the theoretical picture has inspired intensive ARPES measurements extended to the AFM heterostructure members of this family, i.e. MnBi$_4$Te$_7$ consisting of alternating SL and QL sequence and MnBi$_6$Te$_{10}$ consisting of alternating SL and two QLs sequence [44,80]. The enlarged distance between SLs in MnBi$_4$Te$_7$ and MnBi$_6$Te$_{10}$ reduces the AFM exchange interaction. Consequently, MnBi$_4$Te$_7$ has an AFM ground state with $T_N \sim 13\ K$, while MnBi$_6$Te$_{10}$ is AFM below $T_N \sim 10.7\ K$. [81-84]. Due to the vdW interaction between SLs and QLs, MnBi$_4$Te$_7$ has two natural cleavage planes (SL and QL terminations) while MnBi$_6$Te$_{10}$ has three (SL, QL, and double QL terminations). Since the intralayer FM order comes from Mn residing in the central layer of SL, one would expect magnetic gap opening from SL termination and gapless TSS from QL and double QL terminations. However, similar to the case of MnBi$_2$Te$_4$ (0001) surface, all the SL terminations from MnBi$_4$Te$_7$ [49,75,85,86] and MnBi$_6$Te$_{10}$ [75,83,84,87] show (nearly) gapless TSS Dirac cone as presented in Fig. 2b and c. These results suggest that the (nearly) gapless behavior of the TSS Dirac cone is ubiquitous for all the SL terminations of (MnBi$_2$Te$_4$)•(Bi$_2$Te$_3$)$_n$ ($n = 0, 1, 2$).

It is also interesting to look at the TSS behavior from QL and double QL terminations. Both QL terminations from MnBi$_4$Te$_7$ (Fig. 2b, right) and MnBi$_6$Te$_{10}$ (Fig. 2c, middle) present similar features for the TSS. First of all, the TSS Dirac point is buried inside the bulk valence band region as indicated by blue arrows. Secondly, an apparent gap is opened at the upper TSS Dirac cone due to its hybridization with one neighboring bulk valence band. Thirdly, below this hybridization gap, the residual TSS and bulk valance band compose a Rashba-split band (RSB) feature, with the new RSB Dirac point coming from the original TSS Dirac point. Lastly and more importantly, there appears a new band inside the hybridization gap, with its top touching the upper part of the gapped TSS, forming a new gapless Dirac cone. This in-gap state extends from the new Dirac point down to the valence band region, resulting in the generally gapless surface and bulk spectra. The appearance of this new in-gap Dirac cone is intriguing. Based on the above band features, a TSS-RSB hybridization picture has been proposed to explain the complicated band features from both SL and QL terminations [84]. Combining circular dichroism ARPES and first-principles calculations, the existence of RSB and its hybridization with TSS are firmly established by works from several groups [73,79,84,86,88]. The TSS-RSB hybridization can be simulated in a tight-binding simulation. By tuning the hybridization strength, the QL ARPES spectra can be reproduced. According to the simulation, the new in-gap Dirac cone indeed comes from the original bulk RSB (see Fig. 4b-d in ref. [84]). This hybridization picture can also reproduce well the SL ARPES spectra and potentially explain the puzzling gapless behavior of TSS Dirac point, which will be discussed in detail in the following section. For the double QL termination of $MnBi_6Te_{10}$, similar band hybridization features are observed (Fig. 2c, right), but the hybridization gap is too narrow to distinguish any in-gap state.

## 3. Key properties related to the TSS gap

Since its first observation, attempts had been made to explain the gapless behavior of TSS Dirac point from SL in AFM phase with $S$ breaking [47-49]. Before going to the bewildering variety of proposals, we would like to mention the key properties established by various experimental probes, which are closely related to the gapless/gapped behavior of TSS at the SL.

The first one comes from the realization of QAHE observed at $1.4\ K$ based on 5 SLs of $MnBi_2Te_4$ [59], as shown in Fig. 3a. This strongly suggests the existence of 2D Chern insulator state with gapped TSS. By fitting the Arrhenius plot of longitudinal resistance $R_{xx}$ as a function of $1/T$, the energy gap of the thermally activated charge transport can be obtained as $\Delta E = 0.64\ meV$ at zero-field [59]. This energy scale characterizes the minimum energy required to excite an electron from the surface valence to the surface conduction band, two orders of magnitude smaller than the exchange gap predicted [42]. Consequently, the gapless behavior of TSS observed by ARPES may just represent the resolving power of the instrument, and the TSS gap could still exist but be smaller than the energy resolution (typically $> 1\ meV$). The TSS gap size from ARPES measurements varies from being diminished [47,48], to $\sim 10\ meV$ [49] or even larger, with strong sample and spatial dependence [89,90]. The above observations suggest that the gapless behavior of TSS Dirac point at SL termination at AFM phase is unlikely symmetry enforced. It is worth noting that, according to the theoretical definition of AFM TI, the TSS is protected in a weaker sense than the 3D strong TI, as it is generally not stable to disorder [35].

The second key property is the robust A-type AFM order at the surface SL layers as evidenced by measurements using MFM [91] and other techniques. Fig. 3b shows the MFM image taken after field cooling at $0.6\ T$, with the tip polarized by a magnetic field $-0.3\ T$ perpendicular to the sample surface. Clear contrast in the image illustrates several domains, where the magnetic signal changes its sign when crossing the domain walls. If the magnetic moment on the tip is reversed by a magnetic field of $0.3\ T$, all the domains and domain walls change their contrast. Further analysis of the screening effect from fractional QL impurity phases supports the persistence of uniaxial A-type spin order at the top SL layers. One may wonder if the robustness of A-AFM at the top SL layers is also sample dependent. It is thus important to note that other techniques, such as polar reflective magnetic circular dichroism (RMCD) [92] and X-ray magnetic circular/linear dichroism (XMCD/XMLD) [42,45,89,93-95] spectroscopies, also give strong evidence for the existence of net out-of-plane magnetic moments at the sample surfaces.

Even though the robust A-type AFM order is confirmed, what about its influence on the low-energy electronic structure? The (nearly) gapless behavior and its weak temperature dependence across the magnetic transition suggest a negligible effect of the magnetic order on the TSS. However, there are other bands close to the Fermi level which are surprisingly affected by the magnetic order. As exemplified in Fig. 3c by the bulk conduction bands labeled as $CB1_a$ and $CB1_b$, at high temperatures they merge into one band CB1, while their splitting starts when the temperature decreased to $T_N$ and reaches $\sim 35-45\ meV$ below $10\ K$ [48,49,96]. It is further reported that CB2 also shows a Rashba-like feature and band splitting below $T_N$ [76]. These results firmly demonstrated the third key property of MnBi$_2$Te$_4$, i.e. the coupling between AFM order and the low-energy bands.

The fourth one comes from a material point of view concerning the disorders typically present in transition metal chalcogenides. In MnBi$_2$Te$_4$ family, various types of disorders, such as $Bi_{Te}$ antisites (i.e. Bi atoms at the Te sites) located in the surface layer, $Mn_{Bi}$ substitutions (Mn-Bi intermixing) in the second and central atomic layer, and Mn vacancies ($V_{Mn}$) are observed by combining many experimental tools such as scanning transmission electron microscope (STEM) imaging, scanning tunneling microscopy (STM) imaging, single-crystal X-ray diffraction (SCXRD), energy dispersive x-ray (EDX) analysis in scanning electron microscope (SEM) [45,46,74,90,97-105] and even density functional theory (DFT) calculations [106]. We will show in the next section that certain types of disorders may strongly affect the magnetic response of TSS.

### 4. Magnetic reconfiguration to explain the (nearly) gapless TSSs

The (nearly) gapless behavior of TSS and its weak temperature dependence across the AFM order lead to a natural speculation of surface magnetic reconstruction. Deviations from the A-AFMz type (Fig. 4a), such as disordered magnetic structure (Fig. 4b), G-AFM with intralayer and interlayer AFM (Fig. 4c) and AFMx with FM in-plane moments (Fig. 4d), are considered in calculations [47,65,67,107]. As shown in Fig. 4b-d, all the three deviations can lead to gapless TSS. Since it remains as a technical challenge to determine the magnetic structure for the topmost SL, it would be insightful to examine the specific features from ARPES measured spectra corresponding to the various magnetic structure. As shown in the right panel of Fig. 4d, for A-AFMx, the net magnetic moments break the in-plane rotation symmetry and leads to TSS constant energy contour (CEC) with only mirror symmetry. Further consideration of spin texture even shows the mirror symmetry breaking [107]. For G-AFM, the double-sized unit cell in the AFM phase may leads to in-plane band folding feature compared to the paramagnetic phase. For the disordered case, the lack of any oriented moment would retain the sixfold rotation symmetry of TSS CEC, while for the A-AFMz, the net out-of-plane moments in the top SL layer coupled to the TSS will break the sixfold rotation symmetry and leave only threefold rotation symmetry. Further ARPES studies with three components ($P_x, P_y, P_z$) spin resolution and ultrahigh energy, momentum resolution are highly encouraged to distinguish the above features.

Another type of magnetic reconfiguration is the formation of domains and domain walls illustrated in Fig. 4e (left panel [28]). The existence of magnetic domain walls in MnBi$_2$Te$_4$ surface has been observed experimentally [91,108]. Gapless chiral boundary modes are topologically protected to exist in the presence of opposing magnetic domains [109], which is confirmed by the first-principles-based tight-binding model (Fig. 4d, right panel [110]). Note that this edge mode is strictly gapless, while the TSS gap size in MnBi$_2$Te$_4$ shows sample and spatial dependence [89,90]. It is further argued that, as the typical domain size is ~10 μm, the tiny contribution of chiral edge states at domain walls is insufficient to explain the gapless topological surface states [91].

One more sophisticated ferrimagnetic structure has been experimentally observed [100,101] and employed to account for the much reduced TSS gap [97]. As shown in Fig. 4f, the Mn-Bi intermixing could introduce $Mn_{Bi}$ defects in the second and sixth atomic layers counting from the surface, with its moments antiparallel to that of the central Mn layer. Due to the predominant localization of TSS density of states to the Te-Bi-Te layer, the moments of $Mn_{Bi}$ defects counteract that from the central Mn layer, leading to TSS gap reduction. The inhomogeneity of $Mn_{Bi}$ defects could explain the sample and spatial dependence of TSS gap size, suggesting that improving the sample crystalline quality to suppress the Mn-Bi intermixing is a crucial task for the near future study [97]. However, to really correlate the $Mn_{Bi}$ defect density to the TSS gap size, one needs to perform *in situ* ARPES and STM measurements for the same region of the sample surface, which is hardly feasible considering that these two techniques have "field of view" with orders of magnitude difference.

It is also reasonable to check the effective coupling between Mn $d$ orbitals contributing magnetism and Bi/Te $p$ orbital related to the TSS. According to a resonant photoemission study [49], the Mn $3d$ states are mainly located 4 $eV$ below the Fermi level (Fig. 4g), negligible in the energy range where nontrivial topology arises, indicating weak coupling between magnetism and TSS. This weak $p-d$ hybridization scenario seems to contradict the observation of magnetism-induced conduction band splitting as discussed in section III. Based on the above analyses, there remain plenty of challenging experiments to perform, microscopically or spectroscopically, to determine the topmost SL magnetic structure accounting for the (nearly) gapless TSS.

## 5. TSS redistribution to explain its (nearly) gapless behavior

In this section we introduce the TSS redistribution picture which could attribute the (nearly) gapless TSS Dirac point to the extended TSS distribution from the topmost SL to the layers beneath, leading to compromised effective magnetic moments the TSS can feel. The compensation of the effective magnetic moments relies on the fact that the topmost SL and the second SL have antiparallel and comparable moments, meaning that this TSS redistribution picture is only applicable to MnBi$_2$Te$_4$ but not the heterostructure members containing nonmagnetic QLs. However, the potential driving force to redistribute the TSS, such as band hybridization, vdW spacing expansion, or charge/defect effect, maybe generally exist in all the members of this family. In the following, we briefly introduce these mechanisms.

Based on the observation of hybridization between the TSS and a pair of RSBs, A TSS-RSB hybridization picture has been proposed [84] to explain the origin of sophisticated band structure for both QL and SL terminations in this material family (see details in section II). Specifically for SL as shown in Fig. 5a, tight-binding model simulation reveals that the TSS Dirac cone has a bulk origin. This inspires a TSS redistribution picture to account for the lack of magnetic effect on TSS in MnBi$_2$Te$_4$. As schematically illustrated in Fig. 5b, in an *ideal* case the TSS predominantly locates on the topmost SL. In the A-AFM configuration, the effective magnetic moments for the TSS are approximately equal to the net FM moments from one SL, which is large enough to open a sizable TSS Dirac gap as expected. In the *actual* case, the TSS distribution extends to the second SL. The interlayer AFM order results in zero net magnetization for the top two SLs and consequently compensated effective magnetic moments for the TSS. In an extreme situation where the top two SLs equally share 50% of TSS localization, gapless TSS appears regardless of the robust surface A-AFM order and its coupling to the band structure. Based on this TSS redistribution picture, a sizeable TSS magnetic gap can be expected if the magnetic compensation effect is eliminated, say, in a FM ground state. This *expected* case will be discussed in the next section.

Such TSS redistribution picture has been indeed supported by numerous model analysis and calculations based on distinct approaches. Starting from a 3D Hamiltonian for bulk MnBi$_2$Te$_4$ and taking into account the spatial profile of the bulk magnetization, an effective model for the TSS has been derived [111]. This model suggests that the diminished surface gap may be caused by a much smaller and more localized intralayer ferromagnetic order and the fact that the surface states are mainly embedded in the first two SLs from the terminating surface (Fig. 5c). To be specific, by using the envelope function the penetration depth of TSS is calculated as ∼16.2 Å, larger than the thickness of one SL (∼13.7 Å,). This is in agreement with the results obtained by *ab initio* calculations, which present the spatial charge distribution of the TSS Dirac state at equilibrium structure and for vdW spacing between the first and second SL expanded by 15.3% (Fig. 5e from ref. [89]). With increasing vdW spacing, the TSS is found to shift its dominant occupation from the top SL to the second SL, resulting in reduced effective moments. The TSS Dirac gap is found to decrease and vanish at 15.3% expansion. As the vdW spacing modulation is likely to occur in both magnetic and nonmagnetic vdW TIs, a general three-Dirac-fermion approach can be developed [112] to describe the TSS behavior. As shown in Fig. 5f, the three-Dirac-fermion refers to three TSS Dirac cones located at the top surface of the topmost SL/QL ($D_1$), the bottom surface of the topmost SL/QL ($D_2$) and the top surface of the second SL/QL ($D_3$), respectively. Their coupling is tuned by coupling energies $\Delta_{12}$ and $\Delta_{23}$, with the latter being dependent on the topmost interlayer vdW spacing $d$. Remarkably, unexpected gapless TSS Dirac cones are found to arise due to $d$ expansion, when the total Chern number of the system changes by 1 in this expansion process. It should be emphasized that, in this three-Dirac-fermion approach, the gapless point is topologically protected and comes from the competition between the Zeeman coupling and the Dirac fermion coupling. Such vdW spacing expansion may be introduced by mechanical cleavage process [22,113] before ARPES and STM measurements, yet it is missing the direct evidence from atomic layer resolved probes such as STEM.

Excess surface charge is found to affect the distribution of TSS in a top/bottom cone-dependent manner [90]. Obviously, the smallest gap values should be achieved when the top and bottom TSS cones are mostly and independently located in two adjacent SLs, as they experience an exchange field of the opposite sign ($q = -0.045e_0$ in Fig. 5d). Furthermore, from an *ab initio* calculation, cation co-antisites $Mn_{Bi}$ and $Bi_{Mn}$ (extra Bi replacing Mn) can push the TSS charge toward the second SL, the top SL's influence on the magnetic gap is reduced while the second SL influence is enhanced simultaneously (Fig. 5g) [114], resulting in compensated effective magnetic moments and reduce TSS magnetic gap. It is noted that the existence of excess charge and Mn-Bi intermixing defects is well established in this material family.

## 6. Perspectives to open the TSS magnetic gap

In the context of TSS redistribution picture, gapless TSS comes from compensated magnetic moments as a nature of the A-AFM order. Assuming an FM background, no compensation exists no matter how the TSS redistributes. This indeed offers a great chance to realize a sizeable TSS gap through magnetic engineering based on MnBi$_2$Te$_4$. In fact, large amount of Sb substitution in the Bi sites can indeed transform the MnBi$_2$Te$_4$ ground state from AFM to FM or ferrimagnetic order [98,100,115-117]. The TSS band structure study by ARPES, however, is difficult due to the heavy hole doping induced by Sb substitution. It is worth noting that with small amount of Sb doping, we have observed a TSS gap opening in MnBi$_2$Te$_4$ samples which stay at the AFM phase. Surprisingly this TSS gap size is proportional to the doping level and carrier density, allowing a continuous tunability of gap size [118]. However, this TSS gap is independent of the AFM- paramagnetic transition, with the origin of gap remaining to be investigated.

Another way to realize the FM ground state is through heterostructure engineering. As mentioned in section II, in $(MnBi_2Te_4)\bullet(Bi_2Te_3)_n$ ($n = 0, 1, 2, 3$) family the interlayer AFM coupling between the FM SLs can be reduced by QL spacing. In fact, with 3 QLs spacing ($n = 3$), $MnBi_8Te_{13}$ develops a long-range FM order below $T_C = 10.5\ K$ [119]. This provides a valuable chance to realize the magnetic gap in TSS from SL termination. High-quality $MnBi_8Te_{13}$ single crystals are grown and characterized through structural, magnetic, transport and electronic structure studies [120]. Its crystal structure shown in Fig. 6a was obtained from SCXRD and powder XRD refinement. The temperature-dependent anisotropic magnetic susceptibility (Fig. 6b) shows Curie-Weiss (CW) behavior above $150\ K$ (inset) with the characteristic temperature $\theta_{CW} = 12.5\ K$ and $10.5\ K$ for $H \parallel c$ and $H \parallel ab$, respectively. The larger bifurcation between zero-field cooling (ZFC) and field cooling (FC) magnetization and magnetic hysteresis loop (Fig. 6c) indicate an easy axis along the $c$-axis and an Ising-type exchange interaction between adjacent Mn layers. These properties suggest a FM order with an out-of-plane magnetic moment configuration in $MnBi_8Te_{13}$.

The heterostructure lattice of $MnBi_8Te_{13}$ naturally yields four types of terminations by cleaving the single crystal perpendicular to the $c$ axis, namely, SL, QL, double QL and triple QL terminations. A spatial-resolved ARPES with a Laser beam spot around $5\ \mu m$ was employed to resolve the intrinsic surface band structure from these four distinct terminations and the results are shown here in Fig. 6e-h. We start with those 3 nonmagnetic QL terminations and find very similar TSS-RSB hybridization features as we discovered from QL and double QL terminations of $MnBi_6Te_{10}$ (Fig. 2c [84]), while the triple QL termination shows a TSS Dirac cone resembling that of $Bi_2Te_3$. Emphasizing was put on the SL termination and a clear gap can be found at the TSS Dirac point as indicated by the black arrow in Fig. 6e. The gap size is extracted by fitting the energy distribution curves (EDCs) using multiple Lorentzian peaks. The fitting yields TSS Dirac gap $\sim 28\ meV$ at $7\ K$. Furthermore, this gap size exhibits a monotonical decrease with increasing temperature (Fig. 6d) and gap closing at $11\ K$, right above $T_C$, establishing an FM-induced Dirac-point gap in the SL termination. Although TSS gaps have been observed in other members of this materials family and doped TIs, their magnetic origin remains controversial with particularly the lack of clear temperature dependence [47,48,75,84,121]. Consequently, this observation—that a TSS Dirac cone gap decreases monotonically with increasing temperature and closes right at $T_C$, forming a gapless Dirac cone—represent the direct evidence of TSSs gapped by the magnetic order among all known magnetic topological materials. It is still more desirable to realize the magnetic gap of TSSs in $MnBi_2Te_4$ rather than its heterostructure cousins as the latter present uncontrollable terminations with different magnetism from the exfoliation process. The realization of the TSS magnetic gap in FM MnBi$_8$Te$_{13}$ seems to be consistent with the TSS redistribution picture.

We have discussed several possible microscopic mechanisms of TSS in two aspects, i.e., magnetic configuration (such as disordered magnetic structure and A-AFMx with FM in-plane moments) and TSS redistribution (such as TSS-RSB hybridization, vdW spacing expansion, excess surface charge, and cation co-antisite effect). It should be noted that, to reach any of these situations, an energy barrier needs to be overcome due to its deviation from the bulk ground states. Hence, a natural question rises up: why these situations would occur? Does it occasionally exist in Mn-Bi-Te family or some deep mechanism lead to it? From the perspective of energy competition, a more general self-doping scenario in real samples had been proposed as the essential force that may drive to such situations based on Koopmans' theorem [122]. For the gapped TSS, the energy level of the conduction band minimum is higher than that for the case of the gapless Dirac point. According to Koopmans' theorem, an electronic self-doping would naturally save energy for the gapless TSS. Once such energy gain overwhelms the energy barrier of any situation to redistribute the TSS or magnetization, the gapless behavior of TSS emerges. In this sense, the self-doping may be the deep origin of nearly gapless TSS, while the TSS redistribution or magnetic reconfiguration serves as the intermediate. In this view, the emergence of the magnetic gap of MnBi$_8$Te$_{13}$ could also be understood. Instead of locating at the surface SL that contributes to TSS, most of the self-doped electrons enter the bulk QL bands, thus suppressing the gapless transition of TSS [122]. Therefore, to open the TSS magnetic gap it is favorable to recover its charge neutrality via doping or material optimizing.

Despite the remaining puzzles of the microscopic mechanism of TSS, there are currently experimental advances that may be informative. Related to the vdW gap mechanism, point contact tunneling spectroscopy on MnBi$_2$Te$_4$ observed the signature of the TSS gap, which indicates that a moderate pressure on the surface may deduce the vdW gap expansion to restore the effective magnetic moments for TSS [123]. Related to the excess surface charge and antisite effect, improving the crystalline quality is a direct way to eliminate such effects. Recent efforts in growing MnBi$_2$Te$_4$ single crystals using chemical-vapor-transport (CVT) methods [124,125] have reported higher quality samples marked with higher Mn occupancy on the Mn site, slightly higher $Mn_{Bi}$ antisites, smaller carrier concentration, and a Fermi level closer to the Dirac point, yielding highest mobility of $2500\ cm^2V^{-1}s^{-1}$ in MnBi$_2$Te$_4$ devices [124]. ARPES measurement with ultrahigh energy and momentum resolution, as in the previous studies [47-49,118,120], is called for the exploration of spectroscopic signature of coupling between the FM order and TSS, such as magnetic gap opening and sixfold rotation symmetry breaking [112,126]. Quantum transport measurement is highly expected based on CVT single crystals for quantized conductivity at higher temperature.

**ACKNOWLEDGEMENTS**

This work is supported by the National Natural Science Foundation of China (NSFC) (Grants No. 12074163), Guangdong Basic and Applied Basic Research Foundation (Grants No. 2022B1515020046 and No. 2021B1515130007), the Guangdong Innovative and Entrepreneurial Research Team Program (Grant No. 2019ZT08C044), Shenzhen Science and Technology Program (Grant No. KQTD20190929173815000). C.C. acknowledges the assistance of SUSTech Core Research Facilities.

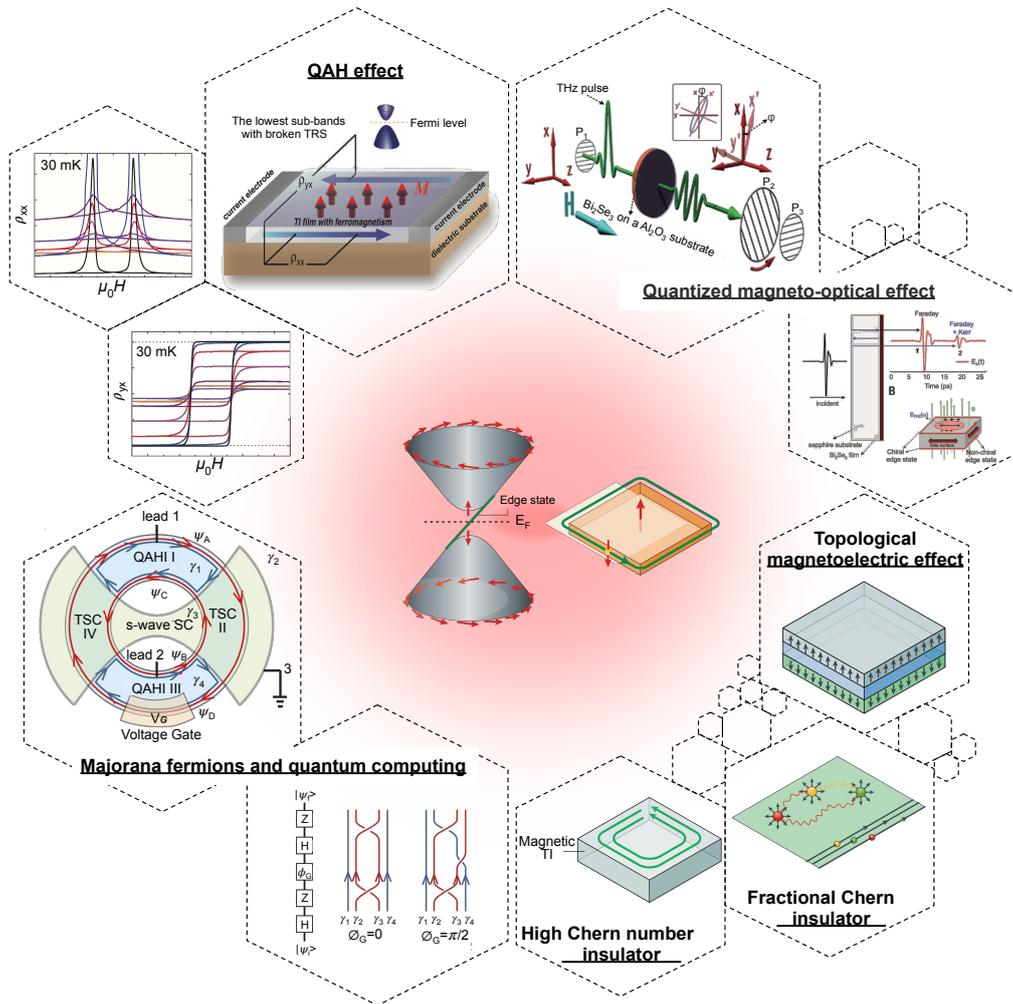

**Figure 1.** Emergent topological phenomena arising from an intrinsic magnetic TI. The interplay of magnetism and band topology offers a great chance to explore the QAHE previously realized on magnetically doped TI films (adapted from [24]) but now potentially at much higher temperatures, the chiral Majorana fermion at the interface of QAHE state and s-wave superconductor which forms the basis of topological quantum computing (from [127]), quantized magneto-optical effect, topological magnetoelectric effect, fractional Chern insulator, high Chern number insulator and so on (from [109]).

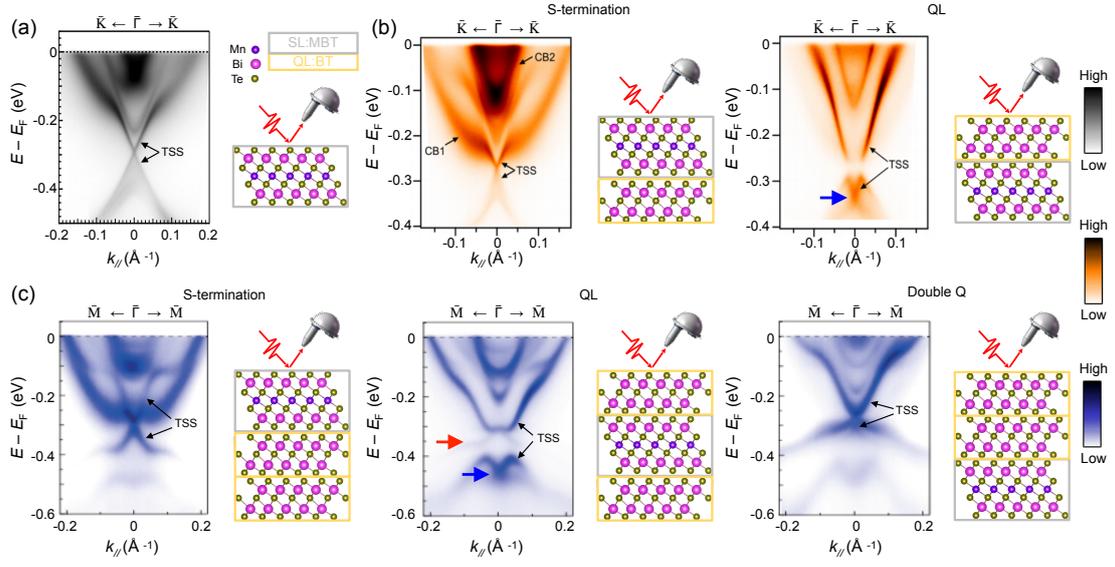

**Figure 2.** Gapless TSSs from all the terminations of (MnBi$_2$Te$_4$)•(Bi$_2$Te$_3$)$_n$ ($n = 0, 1, 2$). (a) Gapless TSS from the (0001) surface of MnBi$_2$Te$_4$, measured at $10\ K$ using photon energy $hv = 6.3\ eV$ [47]. (b) Gapless TSSs from the SL (left) and QL (right) terminations of MnBi$_4$Te$_7$, measured at $11\ K$ using photon energy $hv = 6.3\ eV$ [86]. (c) Gapless TSSs from the SL (left), QL (middle) and double QL (right) terminations of MnBi$_6$Te$_{10}$, measured at $6\ K$ using photon energy $hv = 6.994\ eV$ [75]. Red arrow emphasizes the in-gap states inside the hybridization gap between TSS and bulk valence band while blue arrow indicates the TSS Dirac point.

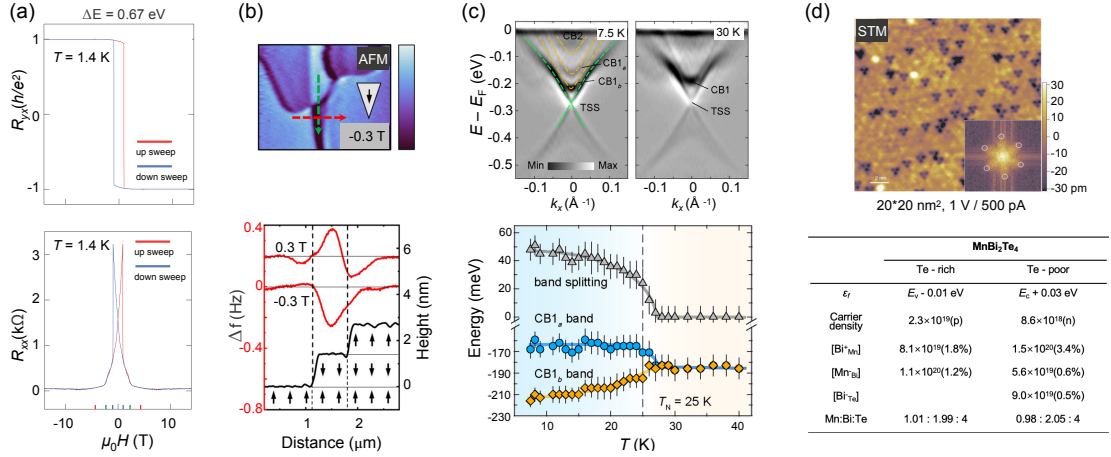

**Figure 3.** Key physical properties related to the gap behaviour of TSS in Mn-Bi-Te family. (a) Realization of QAHE reveals an extremely small TSS gap ($\Delta E = 0.67\ eV$), from [59]. (b) Robust uniaxial A-type AFM order to the surface layers and AFM domains observed using MFM, from [91]. (c) Bulk conduction band splitting related to the AFM order as observed directly in ARPES spectra, from [48]. (d) Top panel shows two types of surface point defects (bright dots and dark triangles, probably corresponding to $Bi_{Te}/Mn_{Te}$ and $Mn_{Bi}$, respectively ) found in atomically resolved topographic image, from [46]; Bottom panel lists the calculated Fermi level ($\varepsilon_f$), free carrier density (and the type of the carrier), as well as densities of the most important intrinsic defects (and concentrations in atomic percent) at both Te-rich and Te-poor limits in $MnBi_2Te_4$, from [106].

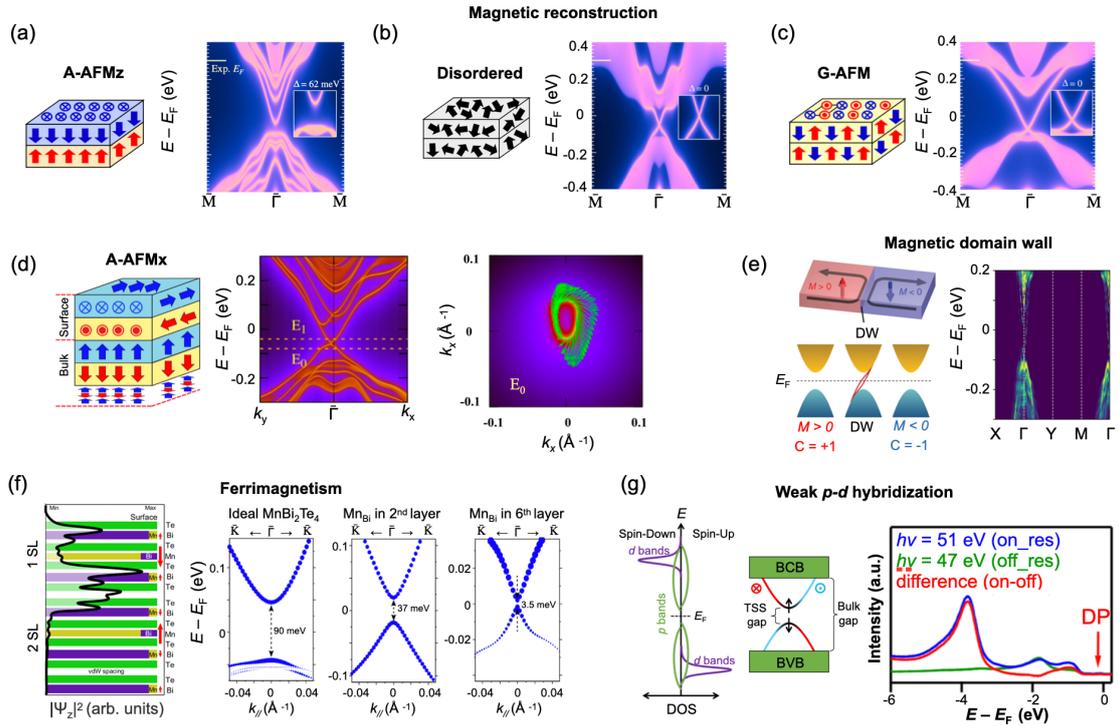

**Figure 4.** Various magnetic reconfigurations accounting for the (nearly) gapless behavior of MnBi$_2$Te$_4$ TSSs. (a)-(c) Prototypical magnetic reconstructions leading to different gap behavior of TSS, including A-type AFM with out-of-plane FM moments (A-AFMz) (a), disordered magnetic moments (b), G-type AFM (c) (Adapted from [47]) and A-type AFM with the magnetic moments along the $x$-axis (A-AFMx) (d) from [107]. (e) Magnetical domain wall edge states (left, from [28]) resemble the gapless TSS, which is further discussed in a first-principles-based tight-binding model (right, from [110]). (f) Native point defects $Mn_{Bi}$ can introduce ferrimagnetism and reduce the TSS gap, from [97]. (g) The hybridization between the Mn $3d$ and Te $5p$ states, as schematically shown in the left two panels (from [69]) is too weak according to the resonant photoemission spectra (right, from [49]) to induce TSS magnetic gap.

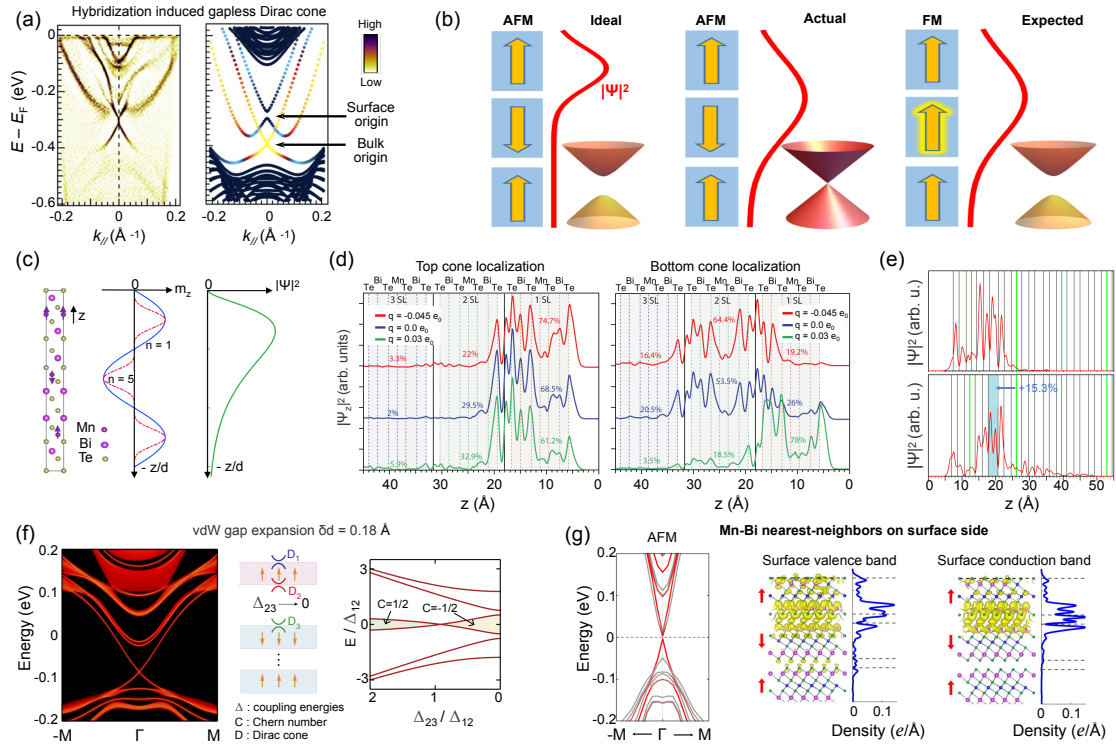

**Figure 5.** TSS redistribution picture explaining the (nearly) gapless behavior of MnBi$_2$Te$_4$ TSSs. (a) ARPES spectra and tight-binding model simulation show the hybridization between TSS and Rashba-split bands, from [84]. (b) Schematic showing the redistribution of TSS density of states and gapped/gapless Dirac cone corresponding to an ideal AFM phase (left), an actual AFM phase (middle) and an expected FM phase (right), from [128]. (c) Distribution of bulk magnetization ($m_z$) and surface state envelope function $\Psi$ according to effective model analysis, from [111]. (d) Surface excess charge induced redistribution of TSS top/bottom cone localization, from [90]. (e) TSS redistribution due to the vdW spacing expansion, from [89]. (f) Three-Dirac-fermion approach to the gapless TSS under vdW spacing expansion, from [112]. (g) Co-antisites (exchanging Mn and Bi atoms in the surface layer) can strongly suppress the magnetic gap down to several $meV$ in the AFM phase, from [114].

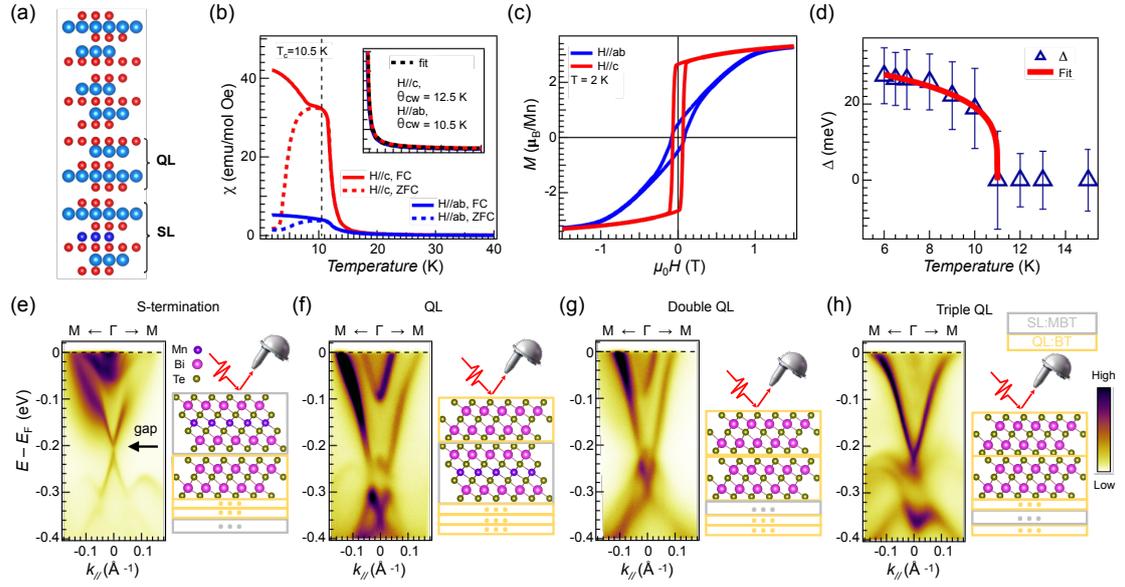

**Figure 6.** Realization of TSS magnetic gap from the S-termination of FM MnBi$_8$Te$_{13}$, from [120]. (a) Schematic crystal structure with one unit of -SL-QL-QL-QL- sequences. (b) magnetic susceptibility shows the FM order with Curie temperature $T_C = 10.5\ K$. (c) Field-dependent magnetization hysteresis at $2\ K$. (d) TSS gap size *vs* temperature shows its FM origin. The gap size is extracted from the TSS of S-termination as shown in (e). (e-h) Termination-dependent band structure measured at $7\ K$ in FM state. The TSS Dirac gap is indicated by a black arrow in (e).